\documentclass[12pt]{article}
\usepackage[top=1in,bottom=1in,left=1.5in,right=1.5in]{geometry}
\usepackage{amsmath,amssymb}
\usepackage{latexsym}% defines $\Box$ for LaTeX2e
  \usepackage[dvips]{pstcol} % PSTricks
\usepackage[dvips]{graphicx}
\newcommand{\N}{\mathbb{N}}
\newcommand{\R}{\mathbb{R}}
\newcommand{\Z}{\mathbb{Z}}

\newcommand{\x}{\mathbf{x}}

\newcommand{\0}{\mathbf{0}}
\newcommand{\1}{\mathbf{1}}

\newcommand{\cM}{\mathcal{M}}

\newcommand{\cS}{\mathcal{S}}
\newcommand{\cT}{\mathcal{T}}

\newcommand{\rQ}{\rm{Q}}
\newcommand{\rS}{\rm{S}}

\newcommand{\lan}{\langle}
\newcommand{\ran}{\rangle}

\newcommand{\an}[1]{\lan#1\ran}
\newcommand{\hs}{\hspace*{\parindent}}
\newcommand{\proof}{\hs \textbf{Proof.\ }}

\newcommand{\trans}{^\top}
\newcommand{\qed}{\hspace*{\fill} $\Box$\\}
\newcommand{\per}{\mathop{\mathrm{perm}}\nolimits}

\newcommand{\pfaf}{\mathop{\mathrm{pfaf}}\nolimits}
\newcommand{\haf}{\mathop{\mathrm{haf}}\nolimits}

\newcommand{\sgn}{\mathrm{sgn}}

\newtheorem{theo}{\bfseries \hs Theorem}[section]

\newtheorem{prop}[theo]{\bfseries \hs Proposition}

\newtheorem{corol}[theo]{\bfseries \hs Corollary}

\numberwithin{equation}{section}
\begin{document}

\title{A polynomial-time approximation algorithm for the number of $k$-matchings in bipartite graphs}

\author{
  Shmuel Friedland and Daniel Levy\\
  Department of Mathematics, Statistics, and Computer Science,\\
  University of Illinois at Chicago\\
  Chicago, Illinois 60607-7045, USA}

 \date{July 28, 2006}

  \maketitle

 \begin{abstract}
 We show that the number of $k$-matching in a given undirected graph
 $G$ is equal to the number of perfect matching of the corresponding graph
 $G_k$ on an even number of vertices divided by a suitable factor.
 If $G$ is bipartite then one can construct a bipartite $G_k$.
 For bipartite graphs this
 result implies that the number of $k$-matching has a polynomial-time approximation
 algorithm.  The above results are extended to permanents and
 hafnians of corresponding matrices.
     \\[\baselineskip] 2000 Mathematics Subject
     Classification: 05A15, 05C70, 68A10.

 \end{abstract}

%-------------------------------------------------------------------------------

\section{Introduction}

 Let $G=(V,E)$ be an undirected graph, (with no self-loops), on the set of
 vertices $V$ and the set of edges $E$.
 A set of edges $M\subseteq E$ is called
 a \emph{matching} if no two distinct edges $e_1,e_2 \in M$ have a
 common vertex.  $M$ is called a $k$-\emph{matching} if $\#M=k$.
 For $k\in\N$ let $\cM_k(G)$ be the set of $k$-matchings in $G$.
 ($\cM_k(G)=\emptyset$ for $k>\lfloor \frac{\#V}{2}\rfloor$.)
 If $\#V=2n$ is even then an $n$-matching is called
 a \emph{perfect matching}.
 $\phi(k,G):=\#\cM_k(G)$ is number of $k$-matchings, and
 let $\phi(0,G):=1$.
 Then $\Phi(x,G):=\sum_{k=0}^{\infty} \phi(k,G)x^k$ is the
 \emph{matching polynomial} of $G$.  It is known that a nonconstant
 matching polynomial of $G$ has only real negative roots \cite{HL}.

 Let $G$ be a bipartite graph, i.e., $V=V_1\cup V_2$ and
 $E\subset V_1\times V_2$.
 In the special case of a bipartite graph where $n=\#V_1=\#V_2$,
 it is well known that $\phi(n,G)$ is given as $\per B(G)$,
 the permanent of the incidence matrix $B(G)$ of the bipartite graph $G$.
 It was shown by Valiant that the computation of the
 permanent of a $(0,1)$ matrix is $\#$\textbf{P}-complete \cite{Val}.
 Hence, it is believed that the computation of the number of perfect
 matching in a general bipartite graph satisfying $\#V_1=\#V_2$ cannot
 be polynomial.

 In a recent paper Jerrum,
 Sinclair and Vigoda gave a \emph{fully-polynomial
 randomized approximation scheme} (\emph{fpras}) to compute the
 permanent of a nonnegative matrix \cite{JSV}.
 (See also Barvinok \cite{Bar}
 for computing the permanents within a simply exponential factor,
 and Friedland, Rider and Zeitouni \cite{FRZ} for
 concentration of permanent estimators
 for certain large positive matrices.)

 \cite{JSV} yields the existence a fpras to compute the number of
 perfect matchings in a general bipartite graph satisfying
 $\#V_1=\#V_2$.
 The aim of this note is to show that there exists fpras to
 compute the number of $k$-matchings for any bipartite graph $G$ and any
 integer $k\in[1,\frac{\#V}{2}]$.
 In particular, the generating matching polynomial of any bipartite
 graph $G$ has a fpras.  This observation can be used
 to find a fast computable approximation to the \emph{pressure} function, as discussed in
 \cite{FP}, for
 certain families of infinite graphs appearing in many models of statistical mechanics,
 like the integer lattice $\Z^d$ .

 More generally, there exists a fpras for computing $\per_k B$, the
 sum of all $k\times k$ subpermanents of an $m\times n$
 matrix $B$, for any nonnegative $B$.  This is done by showing that
 $\per_k B=\frac{\per B_k}{(m-k)!(n-k)!} $
 for a corresponding $(m+n-k)\times (m+n-k)$ matrix $B_k$.

 It is known that for a nonbipartite graph $G$ on $2n$ vertices, the number of
 perfect matchings is given by $\haf A(G)$, the \emph{hafnian} of
 the incidence matrix $A(G)$ of $G$.
 The existence of a fpras for computing the number of perfect
 matching for any undirected graph $G$ on even number of vertices
 is an open problem.  (The probabilistic algorithm suggested in
 \cite{JSV} applies to
 the computation of perfect matchings in $G$, however it is not
 known if this algorithm is fpras.)
 The number of
 $k$-matchings in a graph $G$ is equal to $\haf_k A(G)$, the sum of the hafnians
 of all $2k\times 2k$ principle submatrices of $A(G)$.  We show that
 that for any $m\times m$ matrix $A$ there exists a $(2m-2k)\times
 (2m-2k)$ matrix $A_k$ such that $\haf_k A=\frac{\haf A_k}{(2m-k)!}$.
 Hence the computation of the number of $k$-matching in an arbitrary
 $G$, where $n =O(k)$, has fpras if and only if the number of
 perfect matching in $G$ has fpras.

 \section{The equality $\per_k B=\frac{\per B_k}{(m-k)!(n-k)!} $ }

 Recall that for a square matrix
 $A=[a_{ij}]_{i,j=1}^n \in \R^{n\times n}$, the permanent of $A$ is given as
 $\per A:=
 \sum_{\sigma\in \rS_n} a_{1\sigma(1)}\ldots a_{n\sigma(n)}$,
 where $\rS_n$ is the permutation group on $\an{n}:=\{1,\ldots,n\}$.
 Let $\rQ_{k,m}$ denote the set of all subset of cardinality $k$
 of $\an{m}$.  Identify $\alpha\in \rQ_{k,m}$ with the
 subset $\{\alpha_1,\ldots,\alpha_k\}$ where $1\le
 \alpha_1<\ldots<\alpha_k\le m$.  Given an $m\times n$ matrix
 $B=[b_{ij}]_{i,j=1}^{m,n}\in \R^{m\times n}$ and $\alpha\in
 \rQ_{k,m}, \beta\in \rQ_{l,n}$ we let
 $B[\alpha,\beta]:=[b_{\alpha_i\beta_j}]_{i,j=1}^{k,l}\in
 \R^{k\times l}$ to be the corresponding $k\times l$ submatrix of
 $B$.  Then
 \[\per_k B:=\sum_{\alpha\in\rQ_{k,m},\beta\in\rQ_{k,n}} \per
 B[\alpha,\beta].\]

 Let $G=(V_1\cup V_2,E)$ be a bipartite graph on two classes of
 vertices $V_1$ and $V_2$.  For simplicity of notation we assume
 that $E\subset V_1\times V_2$.  It would be convenient to assume
 that $\# V_1=m,\# V_2=n$.
 So $G$ is presented by $(0,1)$
 matrix $B(G)\in \{0,1\}^{m\times n}$.  That is
 $B(G)=[b_{ij}]_{i,j=1}^{m,n}$ and $b_{ij}=1\iff (i,j)\in E$.
 Let $k\in [1,\min(m,n)]$ be an integer.  Then $k$-matching
 is a choice of $k$ edges in $E_k:=\{e_1,\ldots,e_k\}\subset E$
 such that $E_k$ covers
 $2k$ vertices in $G$.  That is, no two edges in $E_k$ have a
 common vertex.
 It is straightforward to show that $\per_k B(G)$ is the number of
 $k$-matching in $G$.

 More generally, let $B=[b_{ij}]\in\R_+^{m\times n},\;\R_+:=[0,\infty)$ be an $m\times
 n$ nonnegative matrix.
 We associate with $B$ the following
 bipartite graph $G(B)=(V_1(B)\cup V_2(B),E(B))$.
 Identify $V_1(B), V_2(B)$ with $\an{m},\an{n}$
 respectively.  Then for $i\in \an{m},j\in\an{n}$ the edge $(i,j)$
 is in $E(B)$ if and only if $b_{ij}>0$.  Let $G_w:=(V_1(B)\cup V_2(B),E_w(B))$
 be the weighted
 graph corresponding to $B$.  I.e., the weight of the edge $(i,j)\in E(B)$
 is $b_{ij}>0$.  Hence $B(G_w)$,
 the representation matrix of the
 \emph{weighted bipartite graph} $G_w$, is equal to $B$.
 Let $M\in\cM_k(G(B))$.  Then $\prod_{(i,j)\in M} b_{ij}$ is the
 \emph{weight} of the matching $M$ in $G_w$.
 In particular,
 $\per_k B$ is the \emph{total weight} of \emph{weighted} $k$-matchings of $G_w$.
 The weighted matching polynomial corresponding to $B\in \R^{m\times
 n}_+$, or $G_w$ induced by $B$, is defined as:

 $$\Phi(x,B):=\sum_{k=0}^{\min(m,n)} \per_k B \;x^k,\;B\in
 \R_+^{m\times n},\;\per_0 B:=0.$$
 $\Phi(x,B)$ can be viewed as the \emph{grand partition function}
 for the monomer-dimer model in statistical mechanics \cite{HL}.
 (See \S3 for the case of a nonbipartite graph.)
 In particular, all roots of $\Phi(x,B)$ are negative.

 \begin{theo}\label{defBk}  Let
 $B\in \R_+^{m\times n}$ and $k\in
 \an{\min(m,n)}$.  Let

 \noindent
 $B_k\in \R_+^{(m+n-k)\times (m+n-k)}$ be the
 following $2\times 2$ block matrix

 \noindent
 $B_k:=
  \left[
 \begin{array}{cc}B&
 \1_{m,m-k}\\ \1_{n-k,n}&\0
 \end{array}\right]
 $, where $\1_{p,q}$ is a $p\times q$ matrix whose all entries are
 equal to $1$.  Then
 \begin{equation}\label{eqBk}
 \per_k B=\frac{\per B_k}{(m-k)!(n-k)!}.
 \end{equation}

 \end{theo}
 \proof  For simplicity of the exposition we assume that
 $k<\min(m,n)$.  (In the case that $k=\min(m,n)$ then $B_k$ has one
 of the following block structure:
 $1\times 1$, $1\times 2$, $2\times 1$.)
 Let $G_w=(V_1(B)\cup V_2(B),E_w(B)),G_{w,k}=(V_{1}(B_k)\cup V_{2}(B),
 E_{w}(B_k))$ be the weighted graphs corresponding to
 $B,B_k$ respectively.  Note that $G_{w}$ is a weighted subgraph of
 $G_{w,k}$ induced by $V_1(B)=\an{m}\subset \an{m+n-k}=V_1(B_k),
 V_2(B)=\an{n}\subset \an{n+m-k}=V_2(B_k)$.  Furthermore, each
 vertex in $V_1(B_k)\backslash V_1(B)$ is connected exactly to each vertex in
 $V_2(B)$, and each
 vertex in $V_2(B_k)\backslash V_2(B)$ is connected exactly to each vertex in
 $V_1(B)$.  The weights of each of these edges is $1$.  These are
 all edges in $G(B_k)$.  A perfect match in $G(B_k)$ correspond to:
 \begin{itemize}
 \item An $n-k$ match between the set of vertices $V_1(B_k)\backslash V_1(B)$ and the set of
 vertices $\beta'\in \rQ_{n-k,n}$, viewed as a subset of $V_2(B)$.
 \item An $m-k$ match between the set of vertices $V_2(B_k)\backslash V_2(B)$ and the set of
 vertices $\alpha'\in \rQ_{m-k,m}$, viewed as a subset of $V_1(B)$.
 \item A $k$ match between the set of vertices $\alpha:=\an{m}\backslash
 \alpha'\subset V_1(B)$ and $\beta:=\an{n}\backslash
 \beta'\subset V_2(B)$.

 \end{itemize}

 Fix $\alpha\in Q_{k,m}, \beta\in Q_{k,n}$.  Then the total weight
 of $k$-matchings in $G_w(B_k)$ using the set of vertices $\alpha\subset V_1(B_k),
 \beta\subset V_2(B_k)$ is given by $\per B[\alpha,\beta]$.
 The total weight of $n-k$ matchings using $V_1(B_k)\backslash V_1(B)$ and
 $\beta'\subset V_2(B_k)$ is $(n-k)!$.
 The total weight of $m-k$ matchings using $V_2(B_k)\backslash V_2(B)$ and
 $\alpha'\subset V_1(B_k)$ is $(m-k)!$.  Hence the total weight
 of perfect matchings in $G_w(B_k)$, which matches the set of vertices $\alpha\subset V_1(B_k)$
 with the set $\beta\subset V_2(B_k)$ is given by $(m-k)!(n-k)!\per
 B[\alpha,\beta]$.  Thus $\per B_k=(m-k)!(n-k)!\per_k B$.  \qed

 We remark that the special case of Theorem \ref{defBk} where $m=n$
 appears in an equivalent form in \cite{Fr}.

 \begin{prop}\label{equivpm}  The complexity of computing
 the number of $k$-matchings in a bipartite graph
 $G=(V_1\cup V_2,E)$, where

 \noindent
 $\min(\#V_1,\#V_2)\ge k\ge c\max(\#V_1,\#V_2)^{\alpha}$ and $c,\alpha \in (0,1]$,
 is polynomially equivalent to
 the complexity of computing the number of perfect matching in a
 bipartite graph $G'=(V_1'\cup V_2',E')$, where $\#V_1'=\#V_2'$.

 \end{prop}

 \proof  Assume first that $G=(V_1\cup V_2,E), m=\#V_1,n=\#V_2$ and $k\in
 [c\max(\#V_1,\#V_2)^{\alpha},\min(m,n)]$
 are given.  Let $G'=(V_1'\cup V_2',E')$ be the bipartite graph constructed in the
 proof of Theorem \ref{defBk}.   Theorem  \ref{defBk} yields that the number of perfect
 matching in $G'$ determines the number of $k$-matching in $G$.
 Note that $n':=\#V_1'=\#V_2'=O(k^{\frac{1}{\alpha}})$.
 So the $k$-matching problem is a special case of the perfect
 matching problem.

 Assume second that $G'=(V_1'\cup V_2',E')$ is a given bipartite graph
 with $k=\#V_1=\#V_2$.  Let $m,n\ge k$ and denote by $G=(V_1\cup
 V_2, E'),\#V_1=m,\#V_2=n$ the graph obtained from $G$ by adding $m-k,n-k$
 isolated vertices to $V_1',V_2'$ respectively, ($E'=E$).
 Then a perfect matching in $G'$ is a $k$-matching in $G$, and the number
 of perfect matching in $G'$ is equal to the number of $k$-matchings in $G$.
 Furthermore if $k\ge  c\max(m,n)^{\alpha}$ it follows that
 $m,n=O(k^{\frac{1}{\alpha}})$.  \qed

 The results of \cite{JSV} yield.

 \begin{corol}\label{frpk}  Let $B\in \R_+^{m\times n}$ and $k\in
 \an{\min(m,n)}$.  Then there exists a fully-polynomial
 randomized approximation scheme to compute $\per_k B$.
 Furthermore for each $x\in\R$
 there exists a fully-polynomial
 randomized approximation scheme to compute
 the matching polynomial $\Phi(x,B)$.
 \end{corol}

 \section{Hafnians}

 Let $G=(V,E)$ be an undirected graph on $m:=\#V$ vertices.
 Identify $V$ with $\an{m}$.  Let $A(G)=[a_{ij}]_{i,j=1}^m\in \{0,1\}^{m\times
 m}$ be the incidence matrix of $G$, i.e. $a_{ij}=1$
 if and only if  $(i,j)\in
 E$.  Since we assume that $G$ ia undirected and  has no self-loops, $A(G)$ is
 a symmetric $(0,1)$ matrix with a zero
 diagonal.
 Denote by $\rS_{m}(\cT)\supset \rS_{m,0}(\cT)$ the set of symmetric matrices
 and the subset of symmetric matrices with zero
 diagonal respectively, whose nonzero entries are in the set
 $\cT\subseteq \R $.
 Thus any $A=[a_{ij}]\in\rS_{m,0}(\R_+)$ induces
 $G(A)=(V(A),E(A))$, where $V(A)=\an{m}$ and $(i,j)\in E(A)$ if and
 only if $a_{ij}>0$.  Such an $A$ induces a weighted graph $G_w(A)$,
 where the edge $(i,j)\in E(A)$ has the weight $a_{ij}>0$.
 Let $M\in \cM_k(G(A))$ be a $k$-matching in $G(A)$.  Then the weight
 of $M$ in $G_w(A)$ is given by $\prod_{(i,j)\in M} a_{ij}$.

 Assume that $m$ is even, i.e. $m=2n$.  It is well known
 that the number of perfect matchings in $G$ is given by $\haf
 A(G)$, the \emph{hafnian} of $A(G)$.  More general, the total weight
 of all weighted perfect matchings of $G_w(A), A\in \rS_{2n,0}(\R_+)$
 is given by $\haf A$, the hafnian of $A$.

 Recall the definition of the hafnian
 of $2n\times 2n$ real symmetric matrix $A=[a_{ij}]\in\R^{2n\times
 2n}$.  Let $K_{2n}$ be the complete graph on $2n$ vertices,
 and denote by $\cM(K_{2n})$ the set of all perfect matches
 in $K_{2n}$.  Then $\alpha\in \cM(K_{2n})$ can be represented as
 $\displaystyle \alpha^{}_{}=\{(i_1,j_1),(i_2,j_2),..,(i_n,j_n)\} $
 with $ i_k<j_k$ for $k=1,\ldots$.  Denote
 $a_{\alpha}:=\prod_{k=1}^n a_{i_k j_k}$.  Then $\haf
 A:=\sum_{\alpha\in\cM(K_{2n})} a_{\alpha}$.
 Note that $\haf A$ does not depend on the diagonal entries of $A$.
 Hafnian of $A$ is related to the \emph{pfaffian} of the skew symmetric
 matrix $B=[b_{ij}]\in \R^{2n\times 2n}$, where $b_{ij}=a_{ij}$ if
 $i<j$, the same way the permanent of $C\in \R^{n\times n}$ is
 related to the determinant of $C$.  Recall $\pfaf B=\sum_{\alpha\in
 \cM(K_{2n})} \sgn(\alpha) b_{\alpha}$, where $\sgn(\alpha)$ is the
 signature of the permutation $\alpha\in\cS_{2n}$ given by
 $\displaystyle \alpha=\begin{bmatrix}1 & 2 & 3 & 4
 & ..& 2n \\ i_1 & j_1 & i_2 & j_2 &..& j_{n} \end{bmatrix}$.
 Furthermore $\det B=(\pfaf B)^2$.

 Let $A\in \rS_m(\R)$.  Then
 $$ \haf_k A:=\sum_{\alpha\in Q_{2k,m}} \haf A[\alpha,\alpha], \;
 k=1,\ldots,\lfloor \frac{m}{2}\rfloor.$$
 For $A\in \rS_{m,0}(\R_+)$ $\haf_k A$ is the total weight of all
 weighted $k$-matchings in $G_w(A)$.  Let $\haf_0(A):=1$.
 Then the weighted matching polynomial of $G_w(A)$ is given by
 $\Phi(x,A):=\sum_{k=0}^{\lfloor \frac{m}{2}\rfloor} \haf_k A\;x^k$.
 It is known that a nonconstant $\Phi(x,A), A\in \rS_{m,0}(\R_+)$
 has only real negative roots \cite{HL}.

 \begin{theo}\label{defAk}  Let
 $A\in S_{m,0}(\R_+)$ and $k\in
 \an{\lfloor \frac{m}{2}\rfloor}$.  Let
 $A_k\in \rS_{2m-2k,0}(\R_+)$ be the
 following $2\times 2$ block matrix
 $A_k:=
  \left[
 \begin{array}{cc}A&
 \1_{m,m-2k}\\ \1_{m-2k,m}&\0
 \end{array}\right]
 $.  Then
 \begin{equation}\label{eqAk}
 \haf_k A=\frac{\haf A_k}{(m-2k)!}.
 \end{equation}

 \end{theo}

 \proof  It is enough to consider the nontrivial case
 $k<\frac{m}{2}$.  Let $G_w=(V(A),E_w(A)),G_{w,k}=(V(A_k),
 E_{w}(A_k))$ be the weighted graphs corresponding to
 $A,A_k$ respectively.  Note that $G_{w}$ is a weighted subgraph of
 $G_{w,k}$ induced by $V(A)=\an{m}\subset \an{2m-2k}=V(A_k)$.  Furthermore, each
 vertex in $V(A_k)\backslash V(A)$ is connected exactly to each vertex in
 $V(A)$.  The weights of each of these edges is $1$.  These are
 all edges in $G(A_k)$.  A perfect match in $G(A_k)$ correspond to:
 \begin{itemize}
 \item An $m-2k$ match between the set of vertices $V(A_k)\backslash V(A)$
 and the set of
 vertices $\alpha'\in \rQ_{m-2k,m}$, viewed as a subset of $V(A)$.
 \item A $k$ match between the set of vertices $\alpha:=\an{m}\backslash
 \alpha'\subset V(B)$.

 \end{itemize}

 Fix $\alpha\in Q_{2k,m}$.  Then the total weight
 of $k$-matchings in $G_w(A_k)$ using the set of vertices $\alpha\subset V(A_k)$
 is given by $\haf A[\alpha,\alpha]$.
 The total weight of $m-2k$ matchings using $V(A_k)\backslash V(A)$ and
 $V(A)\backslash \alpha$ is $(m-2k)!$.
 Hence the total weight
 of perfect matchings in $G_w(A_k)$, which matches the set of vertices $\alpha\subset V(A_k)$
 is given by $(m-2k)!\haf
 A[\alpha,\alpha]$.  Thus $\haf A_k=(m-2k)!\haf_k A$.  \qed

 It is not known if the computation of the number of perfect
 matching in an arbitrary undirected graph on an even number of vertices,
 or more generally the computation
 of $\haf A$ for an arbitrary $A\in \rS_{2n,0}(\R_+),$
 has a fpras.  The probabilistic algorithm outlined in \cite{JSV}
 carries over to the computation of $\haf A$, however it is an open
 problem if this algorithm is a fpras.
 Theorem \ref{defAk} shows that the computation of $\haf_k A$,
 for $A\in \rS_{m,0}(\R_+)$, has the same complexity as the
 computation of $\haf A$, for $A\in \rS_{2n,0}(\R_+)$.

 \section{Remarks}

 In this section we offer an explanation, using the recent results
 in \cite{FG}, why $\per A$ is a nicer function than $\haf A$.
 Let $A=[a_{ij}]\in \rS_{n}(\R),B=[b_{ij}]\in \R^{n\times n}$.  For
 $\x:=(x_1,\ldots,x_n)\trans \in \R^n$ let
 $$p(\x):=\prod_{i=1}^n (\sum_{j=1}^n b_{ij}x_j), \quad q(\x):=\frac{1}{2}\x\trans
 A\x.$$
 Then $\per B= \frac{\partial^n}{\partial x_1\ldots\partial x_n}
 p(\x)$ and $\haf A= ((\frac{n}{2})!)^{-1}\frac{\partial^n}{\partial x_1\ldots\partial
 x_n} q(\x)^{\frac{n}{2}}$ if $n$ is even.
 Assume that $B\in \R_+^{n\times n}$ has no zero row.
 Then $p(\x)$ is a positive hyperbolic polynomial.  (See the
 definition in \cite{FG}.)  Assume that $A\in \rS_{2m,0}(\R_+)$
 is irreducible.
 Then $q(\x)$, and hence any power $q(\x)^i,i\in\N$, is positive
 hyperbolic if and only if all the eigenvalues of $A$, except the
 Perron-Frobenius eigenvalue, are nonpositive.


\begin{thebibliography}{MMM}

 \bibitem{Bar} A. Barvinok, Polynomial time algorithms to approximate permanents
 and mixed discriminants within a simply exponential factor,
 \emph{Random Structures Algorithms}  14 (1999), 29-61.

 \bibitem{Fr} S. Friedland,  A proof of a generalized van der Waerden conjecture on permanents,
 \emph{Linear Multilin. Algebra} 11 (1982), 107-120.

 \bibitem{FG} S. Friedland and L. Gurvits, Generalized Friedland-Tverberg inequality:
 applications and extensions, \emph{submitted}.

 \bibitem{FP} S. Friedland and U.N. Peled,
 The pressure associated with multidimensional SOFT,
 \emph{in preparation}.


 \bibitem{FRZ} S. Friedland, B. Rider and O. Zeitouni, Concentration of permanent
 estimators for certain large matrices, \emph{Annals of Applied Probability},
 14(2004), 1559-1576.

 \bibitem{HL} O.J. Heilman and E.H. Lieb, Theory of monomer-dimer
 systems, \emph{Comm.\ Math.\ Phys.} 25 (1972), 190--232; Errata
 27 (1972), 166.

 \bibitem{JSV} M. Jerrum, A. Sinclair and E. Vigoda,
 A polynomial-time approximation algorithm for the permanent of a matrix
 with non-negative entries, \emph{J. ACM} 51 (2004), 671-697.

 \bibitem{Val} L.G. Valiant, The complexity of computing the
 permanent, \emph{Theoretical Computer Science} 8 (1979), 189-201.

\end{thebibliography}
\end{document}